\documentclass[reprint,amsmath,amssymb,aps,]{revtex4-1}
\usepackage{graphicx}
\usepackage{dcolumn}
\usepackage{bm}

\newcommand {\be}{\begin{equation}}
\newcommand {\ee}{\end{equation}}
\newcommand {\ba}{\begin{eqnarray}}
\newcommand {\ea}{\end{eqnarray}}
\newcommand {\bea}{\begin{eqnarray}}
\newcommand {\eea}{\end{eqnarray}}

\pdfoutput=1
\begin{document}

\title{Implication of Quadratic Divergences Cancellation in the Two Higgs Doublet Model}
\author{Neda Darvishi} \email{neda.darvishi@fuw.edu.pl}
\author{Maria Krawczyk}
\affiliation{Faculty of Physics, University of Warsaw, Pasteura 5, 02-093 Warsaw, Poland.}

\begin{abstract}
\textbf{Abstract:} {
With the aim of exploring the Higgs sector of the Two Higgs Doublet Model (2HDM), we have chosen the exact and soft $Z_2$ symmetry breaking versions of the 2HDM with non-zero vacuum expectation values for both Higgs doublets (Mixed Model).  We consider two SM-like scenarios: with 125 GeV $h$ and 125 GeV $H$. We have applied the condition for cancellation of quadratic divergences in the type II  2HDM in order to derive masses of the heavy scalars. Solutions of two relevant conditions were found in the considered  SM-like scenarios. After applying the current LHC data for the observed 125 GeV Higgs boson, the precision electroweak data test and lower limits on the mass of $H^+$, the allowed region of parameters shrink strongly.}
\end{abstract}
\maketitle
\section{Introduction}
\label{sec:intro}
The SM describes the physics of elementary particles with a very good accuracy \cite{Glashow}. Experiments have confirmed its predictions with remarkable precision.
One of the most precise aspects of the model is associated with the Higgs sector. However, the SM is not a completely perfect model, since it is unable to provide adequate explanations for many questions in particle and astrophysics 
\cite{Darvishi:2016tni,Darvishi:2016fwo,Bonilla:2014xba,Darvishi:2016gvm,Krawczyk:2015xhl,Darvishi:2017fwr}.
One of the problems of the SM is the naturalness of the Higgs mass. From experimental data, we know that the Higgs boson mass (125 GeV) is of the order of the electroweak scale, but from the naturalness perspective, this mass is much larger than the electroweak scale. This is because of the large radiative corrections to the Higgs mass which implies an unnatural tuning between the tree-level Higgs mass and the radiative corrections. These radiative corrections diverge, showing a quadratic sensitivity to the largest scale in the theory \cite{weinberg}.
Solutions to this hierarchy problem imply new physics beyond the SM, which must be able to compensate these large corrections to the Higgs boson mass. This goal can be obtained with the presence of new symmetries and particles.
 Veltman suggested that the radiative corrections to the scalar mass vanish (or are kept at a manageable level) \cite{Veltman}. This is known as the Veltman condition.

In this paper, we apply Veltman condition to the 2HDM to predict masses of the additional neutral scalars in two possible scenarios, with the SM like-$h$ and the SM like-$H$ bosons. The reader can find similar discussions in \cite{wu, ma, Grzadkowski:2010dn,Chowdhury:2015yja,Biswas:2014uba,Chakraborty:2014oma}.
In 2HDM Lagrangian, four different types of Yukawa interactions arise with no FCNC at tree level \cite{Barger:1989fj,Grossman:1994jb}. In type I, all the fermions couple with the first doublet $\Phi_1$ and none with the $\Phi_2$. In type II, the down-type quark and the charged leptons couple to the first doublet, and the up-type quarks to the second doublet. In type III or the flipped models, the down-type quarks couple to the $\Phi_1$ and the up-type quarks and the charged leptons couple to the $\Phi_2$. In type IV or  the lepton-specific models, all quarks couple to the $\Phi_1$ and the charged leptons couple to the $\Phi_2$.  Among fermions, we include only the dominate top and bottom quark contributions and neglect leptons. Therefore type I and type IV become identical. The same statement is true for the type II and the type III.
In 2HDM type I, vanishing quadratic divergence are possible due to negative scalar quartic couplings, but this solution is contradict to the condition of positivity of the Higgs potential \cite{Chakraborty:2014oma,Ma:2014cfa,Barbieri:2006dq}, therefore we concentrate on the 2HDM Model type II .

\section{Mixed Model with a soft $Z_2$ symmetry breaking}
\label{cqd1}
The Higgs sector of the 2HDM consists of two SU(2) scalar doublets, $\Phi_{1}$ and $\Phi_{2}$.
The 2HDM potential depends on quadratic and quartic parameters, respectively $m_{11}^2, m_{22}^2,m_{12}^2$
and $\lambda_i$ (i= 1..., 5), from which five Higgs boson masses come up after the spontaneous symmetry breakdown (SSB). The most general $SU(2)$ $\times{}$ $U(1)$ invariant Higgs potential for two doublets,
$${\Phi{}}_{1,2}=(\phi_{1,2}^+, \Phi_{1,2}^0)^\dag,$$ 
with a soft $Z_2$ symmetry ($\Phi_1 \to \Phi_1\
, \Phi_2 \to -\Phi_2$) violation is given by:
\begin{widetext}
\begin{align}
 V=&\frac{\lambda_1}{2}(\Phi_1^+\Phi_1)^2+\frac{\lambda_2}{2}(\Phi_2^+\Phi_2)^2+\lambda_3\left(\Phi_1^+\Phi_1\right)\left({\Phi{}}_2^+\Phi_2\right)+\lambda_4\left(\Phi_1^+\Phi_2\right)\left(\Phi_2^+\Phi_1\right)
 \notag
\\&+(\frac{1}{2}\lambda_5({\Phi{}}_1^+{\Phi{}}_2)^2
+h.c.)-\frac{m_{11}^2}{2}\left({\Phi{}}_1^+{\Phi{}}_1\right)-\frac{m_{22}^2}{2}\left({\Phi{}}_2^+{\Phi{}}_2\right)-(\frac{m_{12}^2}{2}(\Phi_1^+\Phi_2)+h.c.).        
\end{align}
\end{widetext}
All parameters 
are assumed to be real, so that CP is conserved in the model.

In order to have a stable minimum, the parameters of the potential need to satisfy the positivity conditions leading to the potential bounded from below.
This behaviour is governed by the quadratic terms, which have the following positivity conditions \cite{Klimenko:1984qx}:
\begin{eqnarray}
 \lambda_1 &>& 0,
 \nonumber \\
 \lambda_2 &>& 0,
 \nonumber \\
 \lambda_3 &>& -\sqrt{\lambda_1\lambda_2},
 \nonumber \\
 |\lambda_5| &<&  \lambda_3 + \lambda_4 + \sqrt{\lambda_1\lambda_2}.
 \, 
\end{eqnarray}

The high energy scattering matrix of the scalar sector at tree level contains only s–wave amplitudes that are described by the quartic part of the potential. The tree level unitarity constraints require that the eigenvalues of this scattering matrix, $|\Lambda_i|$,  be less than the unitarity limit\cite{Plehn:2009nd,Ginzburg:0312374}.  
This means, the requirement $|Re(a_0)| < 1/2$  (or $|(a_0)| < 1$, with $a_0$ being the 0th partial s-wave amplitude for the $2 \to 2$ body scatterings) corresponds to  $|\Lambda_i|\leq 8\pi$.
Finally, one can also impose harder constraints on the parameters of the potential based on arguments of perturbativity, by demanding that the quartic Higgs couplings fulfill $|\lambda_i|\leq 4\pi$ \cite{Plehn:2009nd,Ginzburg:0312374, Swiezewska:2012ej,Krauss:2017xpj,Cacchio:2016qyh,my2}.

The Mixed Model is based on the vacuum with nonzero VEV for both doublets, respectively $<{{\phi{}}_{\ }}_1^0>=\frac{{\upsilon{}}_1}{\sqrt{2}}\neq0$ and 
$<{{\ \phi{}}_{\ }}_2^0>=\frac{{\upsilon{}}_2}{\sqrt{2}}\neq0$, with $\upsilon^2=\upsilon_1^2+\upsilon_2^2 $.
 The minimization conditions are as follows:
\bea
m_{11}^2={\upsilon{}}_1^2{\lambda{}}_1+{\upsilon{}}_2^2({\lambda{}}_{345}-2\nu{}),\\
m_{22}^2={\upsilon{}}_2^2{\lambda{}}_2+{\upsilon{}}_1^2({\lambda{}}_{345}-2\nu{}),
\eea
where ${\lambda{}}_{345}\equiv{}{\lambda{}}_3+{\lambda{}}_4+{\lambda{}}_5$ and
$\nu \equiv m_{12}^2/(2{\upsilon{}}_1{\upsilon{}}_2)$. It is well known that such minimum is the same a global minimum, i.e. vacuum \cite{my2}.

 There are five Higgs particles, with masses as follows: 
\bea
M_{H^{\pm{}}}^2=(\nu{}-\frac{1}{2}\left({\lambda{}}_4+{\lambda{}}_5\right)){\upsilon{}}^2,\\
M_A^2=(\nu{}-\lambda_5 ){\upsilon{}}^2,
\label{MA}
\eea
with the other two mass squared, $M_{h,H}^2$, being  the eigenvalues of the matrix ${\cal {M}}^2$
\bea
{\cal {M}}^2=\left[\begin{array}{
cc}
\cos^2\beta\lambda_1+\sin^2\beta\nu{} &
({\lambda{}}_{345}-\nu)\cos\beta\sin\beta \\
({\lambda{}}_{345}-\nu)\cos\beta\sin\beta &
{\sin^2\beta\lambda{}}_2+\cos^2\beta\nu
\end{array}\right] \upsilon^2, \label{r1} \nonumber \\
\eea
where $\tan\beta{}={\upsilon{}}_2/{\upsilon{}}_1$. This matrix, written 
in terms of the mass squared of physical particles $M_{h,H}^2$, with $M_H \ge
 M_h$, and the mixing angle $\alpha{}$ is given by
 \bea
{\cal M}^2 =\left[\begin{array}{
cc}
M_h^2\sin^2\alpha{}+M_H^2\cos^2\alpha{} &
\left(M_H^2-M_h^2\right)\sin\alpha{}\cos\alpha{} \\
\left(M_H^2-M_h^2\right)\sin\alpha{}\cos\alpha{} &
M_H^2{\sin}^2\alpha{}+M_h^2{\cos}^2\alpha{}
\end{array}\right]. \label{r2} \nonumber \\
\eea


 
The ratio of the coupling constant ($g_i$) of the neutral Higgs boson to the corresponding SM coupling $g_i^{SM}$,
called the relative couplings 
\bea
\centering
\chi_i=\frac{g_i}{g_i^{SM}},
\eea
 are summarised in the table \ref{tab:couplings} (see e.g. reference \cite{CP2005}).
\begin{table*}
 \centering
 \small
 \begin{tabular}{l c c c }
  \hline\hline
   & $\chi_{V}$($W$ and $Z$) & $\chi_{u}$(\text{up-type quarks} ) & $\chi_{d}$(\text{down-type quarks} ) \\
  \hline
   $h$ & $\sin(\beta-\alpha)$ &$ \sin(\beta-\alpha)+\frac{1}{\tan\beta}\cos(\beta-\alpha)$ & $\sin(\beta-\alpha)-\tan\beta\cos(\beta-\alpha)$ \\
 $H$ & $\cos(\beta-\alpha)$ &$ \cos(\beta-\alpha)-\frac{1}{\tan\beta}\sin(\beta-\alpha)$ & $\cos(\beta-\alpha)+\tan\beta\sin(\beta-\alpha)$\\
 $A$ &  $ 0$ & $- i\gamma_5 \cot\beta$ & $- i\gamma_5 \tan\beta $\\
  \hline\hline
 \end{tabular}
 \caption{\label{tab:couplings}
  Tree-level couplings of the neutral Higgs bosons to gauge bosons
  and fermions in 2HDM (II).}
\end{table*}
One sees that all basic couplings can be represented by the couplings to the gauge boson V, $\chi_V=\sin(\beta-\alpha) \,\, (\cos(\beta - \alpha)) $ for $h (H)$ 
and the $\tan \beta$ parameter.

So, we consider two cases which define our SM-like scenarios:

$\bullet$ $M_h \sim 125$ GeV,\,
 $\sin(\beta-\alpha)\sim +1 \,\, (\beta-\alpha=\pi/2)$ \, (SM-like $h$ scenario)
 \vskip .3cm
$\bullet$ $M_H \sim 125$ GeV,
$\cos(\beta-\alpha)\sim +1 \,\, (\beta=\alpha)$  \\
\hspace*{ 4.0cm} $\cos(\beta-\alpha)\sim-1 (\beta-\alpha=\pi)$ \, (SM-like $H_\pm$ scenario).

\vskip 1cm

In both cases, we identify a SM-like Higgs boson with the 125 GeV Higgs particle observed at LHC. Therefore, in the SM-like $h$ scenario, the neutral Higgs partner ($H$) can only be heavier, while in the SM-like $H$ scenario - the partner particle $h$ can only be lighter than $\sim$ 125 GeV. 
 
In this analysis, we apply the positivity conditions and perturbative unitarity condition.
We keep the mixing angles $\alpha$ in the range: $-\pi/2< \alpha < \pi/2$ and $0<\beta<\pi/2$. 

\section{Cancellation of the quadratic divergences}
\label{cqd2}

The cancellation of the quadratic divergences at one-loop applied to 2HDM with Model II for Yukawa interaction leads to a set of two conditions \cite{wu}, namely:
\bea
{6M}_W^2+{3M}_Z^2+{\upsilon{}}^2\left(3{\lambda{}}_1+{2\lambda{}}_3+{\lambda{}}_4\right)=\frac{12}{{\cos}^2\beta{}}{m_D}^2,
 \label{CQD1} \nonumber \\ \\
{6M}_W^2+{3M}_Z^2+{\upsilon{}}^2\left(3{\lambda{}}_2+{2\lambda{}}_3+{\lambda{}}_4\right)=\frac{12}{{\sin}^2\beta{}}{m_U}^2.
\label{CQD2} \nonumber \\
\eea
It should be noted that the Eqs. (\ref{CQD1}) and (\ref{CQD2}) do not depend on the choice of a gauge parameter and the cancellation of the quadratic divergences in the tadpole graphs (\cite{Collins:2006ib}) does not give the additional independent conditions \cite{Blumhofer:1994xv}.
 We include only the dominate top and bottom quarks contributions ($m_D\to m_b$, $m_U\to m_t$). Expressing $\lambda$'s parameters by masses and the mass parameter $m_{12}$,
we have 
\bea
\binom{{\delta{}}_1}{{\delta{}}_2}=\left(\begin{array}{
cc}
A_{11} & A_{12} \\
A_{21} & A_{22}
\end{array}\right)\binom{M_h^2}{M_H^2},\label{cc}
\eea
where
 \begin{widetext}
\bea
{\delta{}}_1=\frac{12m_b^2}{\cos^2\beta{}}-{6M}_W^2-{3M}_Z^2-{2M}_{H^{\pm{}}}^2-M_A^2+\frac{{m}_{12}^2}{2\sin\beta{}\cos\beta{}}[1+3\tan^2\beta{}],
\eea
\bea
{\delta{}}_2=\frac{12m_t^2}{\sin^2\beta{}}-{6M}_W^2-{3M}_Z^2-{2M}_{H^{\pm{}}}^2-M_A^2+\frac{{m}_{12}^2}{2\sin\beta{}\cos\beta{}}[1+3\cot^2\beta{}],
\eea
\end{widetext}
and
\begin{equation}
A_{11}=\frac{3{\sin}^2\alpha{}}{{\cos}^2\beta{}}-\frac{2\sin\alpha{}\cos\alpha{}}{\sin\beta{}\cos\beta{}},
\end{equation}
\begin{equation}
A_{12}=\frac{3{\cos}^2\alpha{}}{{\cos}^2\beta{}}+\frac{2\sin\alpha{}\cos\alpha{}}{\sin\beta{}\cos\beta{}},
\end{equation}
\begin{equation}
A_{21}=\frac{3{\cos}^2\alpha{}}{{\sin}^2\beta{}}-\frac{2\sin\alpha{}\cos\alpha{}}{\sin\beta{}\cos\beta{}},
\end{equation}
\begin{equation}
A_{22}=\frac{3{\sin}^2\alpha{}}{{\sin}^2\beta{}}+\frac{2\sin\alpha{}\cos\alpha{}}{\sin\beta{}\cos\beta{}}.
\end{equation}
In the following section, we solve the Eqs. (\ref{cc}), expressing the condition for cancellation of the quadratic divergences to derive masses of the partner Higgs particles.

\section{Approximate solution of the cancellation conditions}\label{AAa}
\label{cqd3}
It is useful to look first at the approximate solution, which can be obtained analytically.
In the 2HDM model with the soft $Z_2$ symmetry breaking we have for the strict SM-like (alignment) scenarios, with $\sin(\beta-\alpha)=1$ or $\cos(\beta-\alpha)=1$:
\bea
{SM-like}\, \, h: \lambda_1-\lambda_2=(\tan^2 \beta-\frac{1}{\tan^2 \beta })(\frac{M_H^2}{v^2} -\nu), \nonumber
\\
\label{23}
\eea
\bea
{SM-like}\, \, H_+: \lambda_1-\lambda_2=(\tan^2\beta-\frac{1}{\tan^2 \beta }) (\frac{M_h^2}{v^2} -\nu).\nonumber
\\\label{24}
\eea
These formula follow directly from equations (\ref{r1}) and (\ref{r2}). The difference of $\lambda_1$ and $\lambda_2$ is given in terms of $\tan \beta, \nu$ and the mass of the neutral partner for the SM-like $h$ or $H$ particle, mining respectively the $H$ or the $h$ boson.
From difference of equations (\ref{CQD1}) and (\ref{CQD2}) we found that
\begin{eqnarray}
\lambda_1-\lambda_2=\frac{4}{v^2}
(\frac{{m_b}^2}{{\cos^2}\beta}-\frac{{m_t}^2}{{\sin^2}\beta})
 \nonumber \\
 =\frac{4 {m_b}^2}{\upsilon^2}(1-\frac{{m_t}^2/{m_b}^2}{{\tan^2}\beta})&(1+{\tan^2}\beta).\label{18}
 \end{eqnarray}

\begin{figure}
\begin{center}
\includegraphics[width=.5\textwidth]{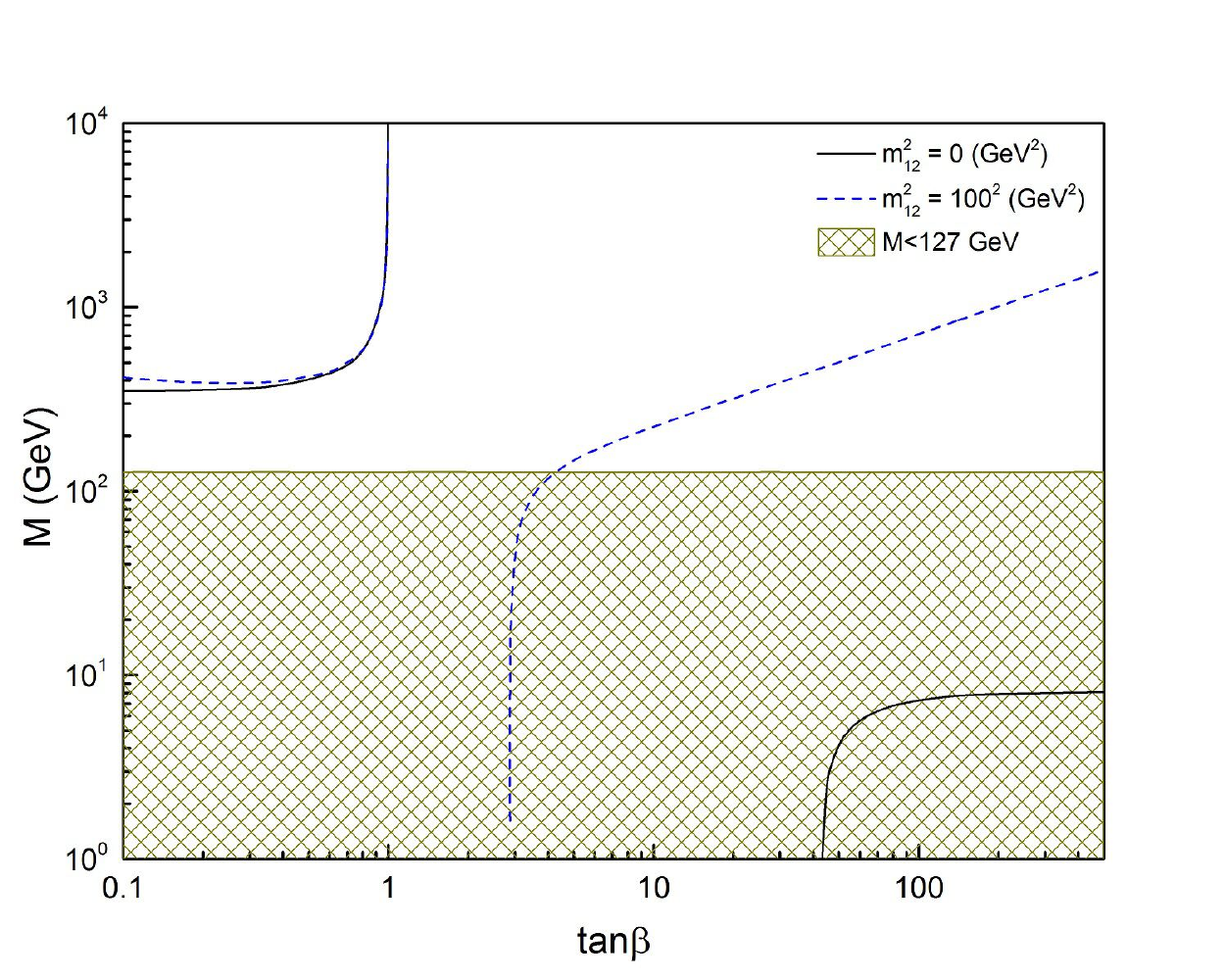}
\caption[h]{The mass of the partner of the SM-like $h(H)$ Higgs particle versus $\tan\beta$ based on Eq. (\ref{M}) for $m_{12}^2=0$ and $m_{12}^2=100^2$ GeV$^2$.}
\label{T2}
\end{center}
\end{figure}

Combining the equations (\ref{23}), (\ref{24}) and (\ref{18}), we obtain the following expressions for masses squared of the partner of the SM-like $h$ or $H$ Higgs particle, 
\bea M^2 =4 m_b^2 \frac{\tan^2 \beta-\frac{m_t^2}{m_b^2}}{\tan^2 \beta-1}+\nu v^2.\label{M}\eea
Obviously, the above prediction for the mass of the partner Higgs particle has been obtained without additional constraints. We have plotted $M$ versus $\tan\beta$, as given by Eq. (\ref{M}) for $m_{12}=0$ and 100 GeV, in the figure \ref{T2}. 
In the figure the mass limit $\sim$ 127 GeV for the SM-like particle is used. The hachure area (i.e. all masses less than 127 GeV) is the allowed region for $M_h$ in the SM-like $H$ scenario and the white area (i.e. all masses higher than 127 GeV) is allowed for $M_H$ in the SM-like $h$ scenario. 
Let us look at the $m_{12}=0=\nu$ case.
It is clear that for the SM-like $h$, solutions exist only for $\tan \beta \le 1$ and the mass of $H$ should be larger than $M_{H0}=2m_t$.
The solutions for the SM-like $H_+$ exist for large $\tan \beta$ ($\tan \beta \ge m_t/m_b \approx 43$) with mass of $h$ below $M_{h0} = 2m_b$.
For positive $\nu$, in the intermediate $\tan\beta$ new regions open up e.g. for $m_{12}^2=100^2$ GeV$^2$. 
For negative $\nu (m_{12}^2)$, all curves are lying below the reference $\nu=0$ curves.

\section{Solving the cancellation conditions}
\label{cqd4}

Here we present the results of numerical solutions of Eqs. (\ref{cc})
 for the SM-like scenarios, as described above.
We apply the positivity and the perturbative unitarity constraints on parameters of the model. We have performed three scans for three considered SM-like scenarios (SM-like $h$, SM-like H+ and SM-like H-) 
with the mass window of the SM-like Higgs 124-127 GeV
and the relative coupling to gauge bosons $\chi_V$ between 0.90 and 1.00, in agreement with the newest LHC data, which are presented in table \ref{g}.

We assume $v$ being bounded to the region $246 \; {\rm GeV} \; <v< \; 247 \; {\rm GeV} $ and using following regions of the parameters of the model:
\bea
M_h < M_H \leq 1000 \;\text{GeV}, \; M_{H^{\pm}} \in [360, 800]\; {\rm GeV} ,\nonumber
\\
 m_{12}^2 \in [-400^2, 400^2]\; {\rm GeV^2}, \; M_{A} \in [130, 700]\; {\rm GeV}.\nonumber
 \\
\eea
Note that very recently the new lower bound on $M_{H^{\pm}}$ has been derived, much higher than the used by us in the scan 360 GeV \cite {oldmisiak}, namely: $M_{H^\pm} > 570 - 800$ GeV \cite{Misiak:2017bgg}.
In the calculations, we use $m_t=172.44$ GeV, $m_b=4.18$ GeV, $M_W=80.38$ GeV, $M_Z=91.18$ GeV \cite{CMS, Beringer}.
Solutions of the Eqs. (\ref{cc}) were found by using $\mathtt{Mathematica}$ and independently by a $\mathtt{C++}$ program, written by us.

Performing our scanning we found no solution for the SM-like $H$ scenario
 in both cases $\cos (\beta-\alpha) \sim \pm 1$, for mass of the charged Higgs boson larger than $360$ GeV \cite{oldmisiak}.

For SM-like $h$ scenario there are solutions only for positive $m_{12}$, in the region 200 - 400 GeV.
Figure \ref{cqd-2} shows the correlation between $m_{12}$ and $M_h$ (panel (a)) and the correlation between $m_{12}$ and $M_H$ (panel (b)). In both panels, the first region from the left (lower $m_{12}$ region) is obtained for large $\tan \beta$ (above 40), while the right one corresponds to low $\tan \beta$ (below 5). 
Figure \ref{cqd-2}(c) shows the correlation of $\tan\beta$ with $M_H$.
In the low $\tan\beta$ the lower limit for $M_H$ is 500 GeV. 
The correlation between $\tan\beta$ versus $M_A$ and $\tan\beta$ versus $M_H^{\pm}$ are similar to correlation between $\tan\beta$ versus $M_H$.
Also, in parts (d), (e) and (f) of the Figure \ref{cqd-2}, we have shown the correlations $M_H$ vs $M_A$, $M_{H^{\pm}} $ vs $M_A$ and 
$M_{H^{\pm}}$ vs $M_H$, respectively.

\begin{figure*}[th]
\centering
\includegraphics[width=0.8\textwidth]{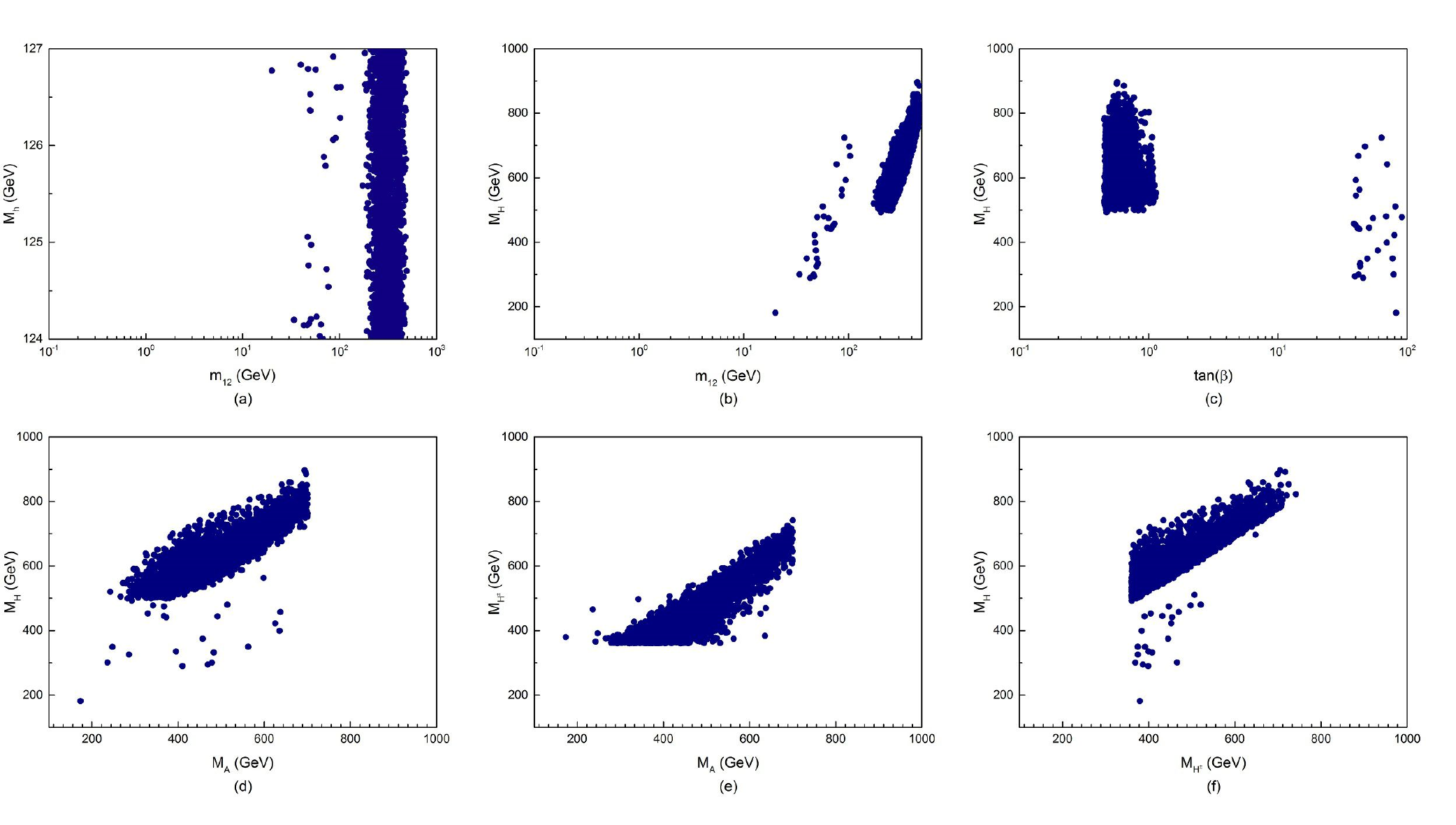} \label{cqd-2}
\caption[]{The correlation between $m_{12}$ and $M_h$ (a), $m_{12}$ and $M_H$ (b), $\tan\beta$ and $M_H$ (c),$M_A$ and $M_H$ (d), $M_A$ and $M_H^{\pm}$ (e), $M_H$ and $M_H^{\pm}$(f).}
\label{cqd-2}
\end{figure*}



Below, we will look closer to the obtained results, by confronting the obtained predictions for observables with experimental data.
We propose 5 benchmarks for the SM-like $h$ scenario, which will be discussed in section \ref{cqd6},
in agreement with theoretical constraints (positivity and perturbative unitarity) and experimental constraints, mainly coming from the measurements of SM-like Higgs boson. In addition, properties of the partner particles ($H$) is checked, which is important for future search. 
In such calculations the $\mathtt{2HDMC}$ program was used \cite{2hdmc}.  

\section{Experimental constraints}
\label{cqd5}

We apply experimental limits from LHC for the SM-like Higgs particle $h$ and solve the cancellation conditions by scanning over $M_h$ and the mixing parameters ($\alpha, \beta$), keeping the values of mass and coupling to the gauge bosons (i.e. $\sin(\beta-\alpha)$) within the experimental bounds. We confront the resulting solutions with existing data for the 125 GeV Higgs boson, in particular the experimental data on Higgs boson couplings ($\chi_V$, $\chi_{t}$ and $\chi_{ b}$) and Higgs signal strength ($R_{ZZ}$, $R_{\gamma\gamma}$ and $R_{Z\gamma}$) from ATLAS \cite{Aad:2014eha} and CMS \cite{Khachatryan:2014ira}, as well as the combined ATLAS+CMS results \cite{201606}. Also the total Higgs decay width measured at the LHC is an important constraint, see tables \ref{g} and \ref{stu3}.

We also keep in mind other existing limits on additional Higgs particles which appear in 2HDM, as the lower mass limit of the $H^+$ taken to be 360 GeV (based on the earlier analysis \cite{oldmisiak}) and check if the obtained solutions are in agreement with
the oblique parameters $S$, $T$, $U$ constraints, being sensitive to presence of extra (heavy) Higgses that are contained in the 2HDM. 

Below, the following short notation will be used:
$t_\beta$, $ s_{\beta-\alpha}$ and $ c_{\beta-\alpha}$ for $\tan\beta$, $ \sin(\beta-\alpha)$ and $\cos(\beta-\alpha)$, respectively.

\section{Benchmarks} 
\label{cqd6}

Results from scan lead to five benchmark points h1-h5, presented in table \ref{g}. 
The results of the scanning show that 
for the SM-like $h$ scenario,
 solutions in agreement with existing data only exist for small $\tan\beta$ (0.45 -$1.07$). 
Values of observables for the SM-like Higgs particle $h$ for five benchmark points h1-h5 are presented in table \ref{stu3}.
The large $\tan\beta$ solutions, (above $42$) exist, however they lead to too large $R_{\gamma \gamma}$, (above 2). 
For convenience, we add the experimental data to both tables (with 1 $\sigma$ accuracy from the fit assuming $|\chi_v|\le 1$ and $B_{SM} \ge 0$). 

Benchmarks correspond to solutions with masses $M_H \sim 505 - 827$ GeV
and $M_A \sim 270 - 650$ GeV and $M_H^{\pm{}} \sim 375 - 646 $ GeV. The newest result of the reference \cite{Misiak:2017bgg} with lower bound on $M_{H^{\pm}} \sim 570 - 800$ GeV can limit our benchmarks to (h3, h4) only.
 There is a small tension, at the 2 $\sigma$, for all benchmarks for the coupling $hb\bar{b}$ with the newest combined LHC result \cite{201606}, which has surprisingly small uncertainty 0.16 (before the individual results were ATLAS $0.61^{+0.24}_{-0.26}$  \cite{Aad:2014eha} and CMS $0.49^{+0.26}_{-0.19}$ \cite{Khachatryan:2014ira}, in perfect agreement with all our benchmarks). Also, benchmarks (h2,h3) correspond to slightly too small mass of the $h$ in the light of the new combined CMS and ATLAS value of 125.09 $\pm$ 0.24 GeV \cite{201606m}. 

\begin{figure}[th]
\centering
\includegraphics[width=0.45\textwidth]{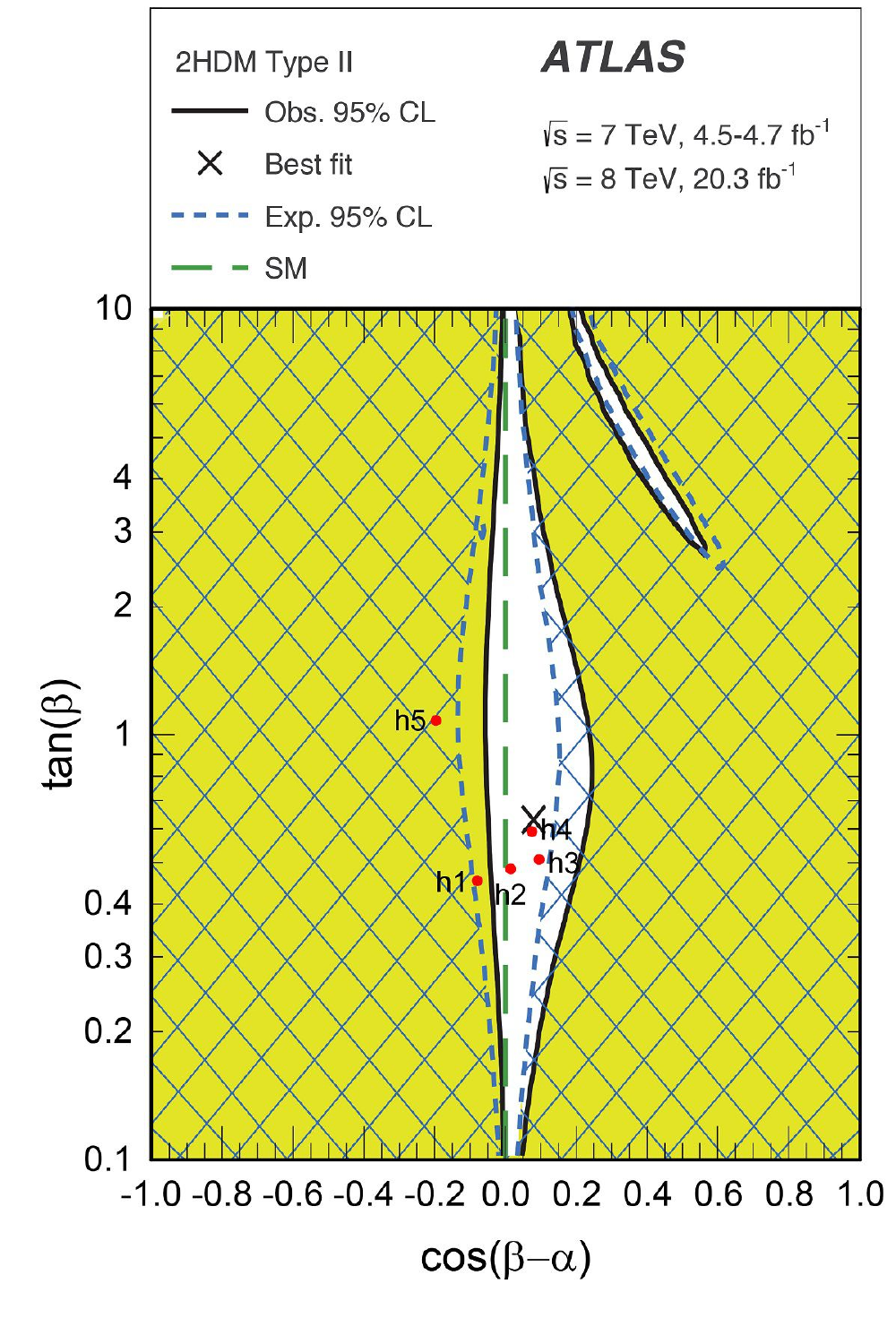} 
\caption{Regions of the 2HDM Model II excluded by to the measured rates of Higgs boson production and
decays by  ATLAS \cite{cite-a-sz} and our benchmarks points h1-h5. The white allowed region contains the best fit point denoted by a cross. Solid (short dashed) lines correspond to the observed (expected) 95 $\%$ limits, the long-dashed vertical line presents a SM prediction. }
\label{fig3}
\end{figure}

We compare our benchmarks to the experimental data on 2HDM (II) on the plot $\tan \beta$ versus $\cos(\beta-\alpha)$, see figure \ref{fig3}. Benchmark h4
is very close to the best fit point found by ATLAS. We would like to point out that our benchmarks results from the cancellation of the quadratic divergences and no fitting procedure has been performed. Note, that the h4 benchmark corresponds to heavy and degenerate $A$ and $H^+$ bosons, with mass $\sim$ 650 GeV, while $H$ is even heavier with mass $\sim$ 830 GeV.

\begin{table*}
\scalebox{0.8}
{\begin{tabular}{|||c|c|c|c|c|c|c|c|c|c|}
\hline 
B mark&$\alpha$ & $t_\beta$ & $s_{\beta-\alpha}$ & $c_{\beta-\alpha}$ & $M_h$ & $M_H$ & $M_A$ & $M_H^{\pm{}}$ & $m_{12}^2$ \\
\hline
\hline
\hline
exp & - &- &1.00 (0.92-1.00) &- & $125.09\pm0.24$ &-&-&-& - \\
\hline
h1&-1.24627&	0.451897&	0.995014&   -0.0997376&	124.426	&573.832&	444.16	&454.424  & $(281.0690)^2$ \\ 
\hline 
h2&-1.10678&	0.481736&	0.999886&   0.0150858&	124.082	&505.298&	266.59	&375.488	& $(191.9640)^2$ \\ 
\hline 
h3&-1.00657& 	0.507350& 	0.995518& 	0.0945748& 	124.242	&736.961& 	567.37	&598.392	& $(352.9160)^2$ \\ 
\hline
h4&-0.96384&	0.589698&	0.997252&	0.0740784&	125.252	&826.947&	650.08	&645.560	& $(421.8410)^2$ \\ 
\hline 
h5&-0.94625&	1.077030&	0.980477&   -0.1966340&	125.771	&605.931&	448.48	&438.628	& $(309.6770)^2$ \\ 
\hline
\hline
\hline
\hline
\end{tabular}}
\caption{SM-like $h$ for $\sin (\beta-\alpha) \sim 1$, Higgs bosons masses (in GeV) are shown for various values of angles $\alpha$ and $\beta$. The experimental data for $M_h$ and $s_{\beta-\alpha}$ are from \cite{201606m} and \cite{201606}, respectively. }
\label{g}
\end{table*}
\begin{table*}
\scalebox{0.8}
{\begin{tabular}{|||c|c|c|||c|c|c|||c|c|c|c|}
\hline 
 B point&$ \chi_t^h$ & $\chi_b^h$&$R^h_{\gamma\gamma}$& $R^h_{Z\gamma} $& $\frac{\Gamma_h^{tot}}{\Gamma_h^{{tot}_{SM}}}$&$S$ & $T$ & $U$ \\ 
\hline
\hline
exp & $1.43^{+0.23}_{-0.22}$& $0.57\pm0.16$&$1.14^{+0.19}_{-0.18}$&$< 9$ & $10^{+14}_{-10}/{4.1}$& $0.05\pm0.11$&$0.09\pm0.13$&$0.01\pm0.11$\\
\hline
 h1& 0.77&	1.04& 1.03 & 0.98  &  0.96 &-0.00& -0.02 & -0.00 \\
\hline
 h2& 1.03&	0.99& 1.09 & 0.97  & 0.96 &-0.00& -0.24 & -0.00 \\
\hline
 h3& 1.17&	0.94& 1.05 & 0.98  & 0.95 &-0.00& -0.07 & -0.00 \\
\hline 
 h4& 1.12&	0.95& 1.11 & 1.08 & 0.97 &-0.00& 0.01 & -0.00\\
\hline
 h5& 0.79&	1.19& 1.31 & 1.35  & 0.77 & 0.01& 0.01 & -0.00 \\
\hline
\hline
\hline
\end{tabular}} 
\caption{SM-like $h$, $ \sin (\beta-\alpha) \sim 1$. The $h$ relative couplings, decays rates and $S$, $T$ and $U$. The experimental data for $ \chi_t^h$, $\chi_b^h$,$R^h_{\gamma\gamma}$ and $R^h_{Z\gamma} $ from \cite{201606}, for $\frac{\Gamma_h^{tot}}{\Gamma_h^{{tot}_{SM}}}$ from \cite{tot-w} and for oblique parameters from \cite{Baak:2014ora} are presented. }
\label{stu3}
\end{table*}
It is worth to look for the properties of the heavy neutral Higgs boson $H$, the partner of the SM-like $h$ bosons. The corresponding observables are given in table \ref{Th3}, here the relative couplings are calculated in respect to couplings the would-be SM Higgs boson with the same mass as $H$.
The coupling to $t$ quark is negative, what is easy to understand looking at the table \ref{tab:couplings}. Its absolute value $|\chi^H_V|$ is enhanced,
 as compared to the SM value, and the corresponding ratio varies from 1.1 to 2.30. One observes a huge enhancement in $R_{\gamma \gamma}$, it is from 50 to 153 times larger the SM one, at the same time $Z \gamma $ decay channel looks modest (0.31-1.44). Also, the total width is similar to the one predicted by the SM - the corresponding ratio varies from 0.3 to 1.09. For a possible search for such particle,
the $\gamma \gamma$ channel would be the best. The $ZZ$ channel is hopeless, but the $H$ decays to $ZA$ and $H^+W^-$, govern by $\sin(\beta-\alpha)$ coupling, may be useful. Note, that all heavy Higgs bosons have masses below 850 GeV, in the energy range being currently probed by the LHC. 

\begin{table*}
\scalebox{0.8}
{\begin{tabular}{|||c|c|c|c|||c|c|c|}
\hline 
 B point&$M_H$&$ \chi_t^H$ & $\chi_b^H$&$R^H_{\gamma\gamma}$& $R^H_{Z\gamma}$& $\frac{\Gamma_H^{tot}}{\Gamma_H^{{tot}_{SM}}}$ \\
\hline
\hline
h1 &573.832 &-2.30&0.34 & 69.14& 0.62 &0.88\\
\hline
h2 &505.298 &-2.06&0.49 & 14.88& 0.31 &1.09 \\
\hline
h3 &736.961 &-1.86&0.59 & 152.71& 1.21 &0.78\\
\hline
h4 &826.947 &-1.61&0.66 & 49.95& 1.44 & 0.53 \\
\hline
h5 &605.931 &-1.10&0.85 & 72.79 & 0.79 &0.28 \\ 
\hline
\hline
\hline
\hline
\end{tabular}} 
\caption{SM-like $h$, $ \sin (\beta-\alpha) \sim 1$. The partner $H$ bosons masses, relative couplings and decays rates are given.}
\begin{center}
\label{Th3}
\end{center} 
\end{table*}

\section{Summary and Conclusion}
\label{cqd7}
In this paper, we have investigated the cancellation of the quadratic divergences in the 2HDM, applying positivity conditions and perturbativity constraint. We have chosen the soft $Z_2$ symmetry breaking version of the 2HDM with non-zero vacuum expectation values for both Higgs doublets (Mixed Model) and considered two SM-like scenarios, with 125 GeV $h$ and 125 GeV $H$. 

We have chosen 5 benchmarks in agreements with experimental data for the 125 GeV Higgs particle from the LHC
and checked that our benchmark points are in agreement with the oblique parameters $S$, $T$ and $U$, at the 3 $\sigma$.

We compare our benchmarks to the experimental constraints for 2HDM (II) on the $\tan \beta$ versus $\cos(\beta-\alpha)$ plane. All benchmarks are close to or within the allowed 95$\%$ CL region, especially our benchmark h4 is very close to the best fit point found by ATLAS. We would like to point out that our benchmarks results from the cancellation of the quadratic divergences and no fitting procedure has been performed. Note, that the h4 benchmark corresponds to heavy and degenerate $A$ and $H^+$bosons, with mass $\sim$ 650 GeV, while $H$ is even heavier with mass $\sim$ 830 GeV.
  

\begin{acknowledgments} 
This work is supported in part by the National Science Center, Poland, the
HARMONIA project under contract UMO-2015/18/M/ST2/00518.

ND thanks Prof. M. Misiak and Dr. M.R. Masouminia for their helpful discussions. 
\end{acknowledgments}

\end{document}